\documentclass[default,iicol]{sn-jnl}

\usepackage{graphicx}%
\usepackage{multirow}%
\usepackage{amsmath,amssymb,amsfonts}%
\usepackage{url}
\usepackage{amsthm}%
\usepackage{mathrsfs}%
\usepackage[title]{appendix}%
\usepackage{xcolor}%
\usepackage{textcomp}%
\usepackage{manyfoot}%
\usepackage{booktabs}%
\usepackage{listings}%

\usepackage{float}
\restylefloat{figure}
\restylefloat{table}

 \usepackage{natbib,stfloats}
\usepackage{lmodern}
\usepackage{anyfontsize}

\begin{document}

\title {Analytical modeling of the gravitational potential of irregularly shaped celestial bodies considering three distinct internal structures: application to (21) Lutetia}


\author*[]{\fnm{Marcelo} \sur{L. Mota$^{1}$}}\email{prof.mlmota@ifsp.edu.br}

\author[]{\fnm{Safwan} \sur{A.$^{2}$}}
\equalcont{These authors contributed equally to this work.}

\author[]{\fnm{Antonio} \sur{F. B. A. Prado$^{2}$}}
\equalcont{These authors contributed equally to this work.}

\affil*[1]{Federal Institute  of S\~{a}o Paulo, IFSP, Hortol\^{a}ndia, SP, Brasil.}%
\affil[2]{National Institute for Space Research, S\~{a}o Jos\'{e} dos Campos, SP, Brazil.}


\abstract{
    The classical polyhedral model is one of the most accurate 
    methods currently used to represent the gravitational field 
    of irregularly shaped bodies. However, it assumes a 
    homogeneous density distribution, which may not accurately 
    reflect the internal composition of real objects. This 
    study aims to analyze the effects of the internal structure 
    of asteroid (21) Lutetia on gravitational potential 
    modeling by considering a three-layered composition with 
    distinct densities.\
    
    The gravitational approach adopted in this study is the 
    Potential Series Expansion Method (PSEM), represents models 
    the body as a polyhedron and decomposes it into tetrahedral 
    elements to estimate of the total potential around the 
    asteroid. This estimation involves summing the 
    contributions of each tetrahedron using a direct triple 
    integral over its volume.\\
    Although this method does not achieve the same level of 
    accuracy as the classical polyhedral approach, it offers a 
    reasonable degree of precision, expresses the potential in 
    analytical form, significantly reduces computational time, 
    and, due to the simplified algebraic manipulation of the 
    potential, facilitates the analysis of the asteroid’s 
    internal structural composition.
}

\maketitle

\section{Introduction}\label{Introduction}

   When studying the dynamical properties of a spacecraft’s 
   orbital motion around an asteroid, a primary challenge in 
   mission design is developing a mathematical model that 
   accurately represents the distribution of the gravitational 
   field outside the asteroid. Due to the irregular mass 
   distribution of asteroids, the gravitational force is non-
   central. Numerous studies have addressed this issue 
   (\citet{Werner_1994}; \citet{1994Icar..110..225S}; 
   \citet{1996Icar..121...67S}; \citet{Werner_1996}; 
   \citet{SCHEERES1998649}; \citet{1998Icar..132...53S}; 
   \citet{1999EP&S...51.1173R};  \citet{2002PhDT........85H};  
   \citet{venditti_2013}; \citet{2017MNRAS.464.3552A}; 
   \citet{Mota_2017}; \citet{Mota_2019}; 
   \citet{2023EPJST.232.2961M}). Typically, an asteroid’s 
   gravitational potential is estimated based on its shape, 
   assuming a homogeneous density distribution. However, this 
   remains an approximation, as actual bodies are affected by 
   internal density variations. Therefore, it is important to 
   examine how different mass distributions influence an 
   object’s gravitational field and, consequently, its orbital 
   environment.\\   
    For example, several studies have modeled the gravitational 
    potential of asteroids Ceres and Vesta using spherical 
    harmonic expansions under various internal structure 
    scenarios (\citet{2010LPI....41.2289T}; \citet{2011SSRv..163..461K}; \citet{2014GeoRL..41.1452K}; \citet{2014Icar..240..118P}; \citet{2017MNRAS.464.3552A}). On the other hand, the polyhedral model proposed by \citet{Werner_1996} appears more suitable for evaluating gravitational forces near the surface. It is worth noting that these approaches involve high computational costs in calculating the required integrals. For instance, \citet{2015MNRAS.450.3742C}  applied the mascon gravity framework using a polyhedral model by dividing each tetrahedron into up to three parts. This opens the possibility of incorporating layered internal structures into the calculation of the gravitational potential.\
    
    In this study, we adopt the Potential Series Expansion Method (PSEM)  \citet{Mota_2017} to model the gravitational potential. This approach models the body as a polyhedron, decomposes it into tetrahedral elements, and determines the total potential around the asteroid by summing the contributions of each tetrahedron. This is achieved through direct triple integration using the method of \citet{Lien_1984}, computed for each tetrahedral volume via tetrahedral isometry. PSEM provides reasonable accuracy compared to the classical polyhedral method, expresses the potential analytically, significantly reduces computational time, and simplifies the analysis of the asteroid's internal structure due to its easier algebraic manipulation. This facilitates the study of asteroids composed of multiple internal layers. However, the main drawback of this method is that the potential is expressed as a series, meaning its validity is limited to its convergence domain—specifically, within the Brillouin sphere. Despite this limitation, we have obtained satisfactory results when modeling other asteroids (\citet{Mota_2019}; \citet{2023EPJST.232.2961M}).\
    
    The astronomer Hermann Goldschmidt discovered in 1852 that asteroid (21) Lutetia is in the Asteroid Belt, a region between the orbits of Mars and Jupiter. Based on studies of its surface composition and temperature, \citet{2011Sci...334..492C} concluded that Lutetia is likely formed during the early stages of the Solar System. Furthermore, measurements taken by the European Space Agency’s Rosetta mission revealed that this body has an unusually high density for an asteroid (3.4 g/cm$^{3}$), suggesting that it may be a partially differentiated object with a dense, metal-rich core (\citet{2011epsc.conf.1184P}; \citet{2012P&SS...66..137W}). For these reasons, (21) Lutetia is considered a suitable object for investigating the effects of layered internal structures on its gravitational potential.\\
    This study therefore aims to compute the gravitational 
    field associated with asteroid (21) Lutetia, considering a model with distinct density layers. Accordingly, we determine the asteroid’s equilibrium points, under this non-homogeneous model. Equilibrium points, also known as Lagrangian points, are critical in the study of asteroids and other celestial bodies for several reasons. These points represent positions in space where an object can remain in a stable configuration. Understanding equilibrium points is essential for spacecraft trajectory planning as missions that explore asteroids often rely on these points to position satellites or guide spacecraft into specific orbits. In the context of future asteroid mining, equilibrium points may serve as strategic locations for establishing bases or platforms to extract resources, providing a stable environment with reduced fuel requirements for station-keeping.\\
    To achieve these objectives, Section 2 presents the 
    physical properties of the homogeneous polyhedral shape of (21) Lutetia. Section 3 provides a brief overview of the the
 Potential Series Expansion Method (PSEM) \citet{Mota_2017} and applies it to a three-layered internal structure model. Additionally, we determine the critical points and their stability, compute the Jacobi integral, and derive the zero-velocity surfaces and particular solutions of the system. Finally, the key findings of our study are presented in Section 4.   
\section{Physical properties of the asteroid (21) Lutetia with uniform density}\label{physical_properties}
   \citet{2011Sci...334..487S} modeled the global shape of (21) Lutetia by combining two techniques: stereophotoclinometry (\citet{2008M&PS...43.1049G}) using images obtained by OSIRIS and the inversion of a set of 50 photometric light curves and adaptive optics image contours (\citet{2010A&A...523A..94C}; \citet{article}). Twelve different shape model solutions are listed in the Planetary Data System (PDS\footnote{\href{http://astro.troja.mff.cuni.cz/projects/damit}{https://sbn.psi.edu/pds/}}).
   
   In this study, we selected the non-convex polyhedral shape model of asteroid (21) Lutetia, which consists of 2,962 faces and is available in the PDS database. The body is aligned with its principal axes of inertia, such that the inertia tensor is represented as a diagonal matrix. Consequently, the x-axis is aligned with the smallest moment of inertia (i.e., the longest axis), the z-axis with the largest moment of inertia (i.e., the shortest axis), and the y-axis corresponds to the intermediate moment. The rotation rate of (21) Lutetia is assumed to be uniform around its maximum moment of inertia (z-axis), with a period of 8.168270±0.000001 hours (\citet{2010A&A...523A..94C}). Using the method proposed by \citet{Lien_1984} for integral calculations, Table 1 presents the main physical properties of the selected polyhedral model.


    \begin{table}[!htp]
        \caption{The physical properties of polyhedral models of the asteroid (21) Lutetia.} \label{table_physical_properties}
        \resizebox{0.5\textwidth}{!}{
            \begin{tabular}{ll}
                \hline
                Number of faces                                   & \multicolumn{1}{c}{2926}  \\ \hline
                Density (g.cm$^{-3}$)                                   &  3.4          \\
                Effective diameters (km)                           &  98.1544711     \\
                Areas estimation ($\times 10^4$ km$^2$)            &  3.28457671     \\
                Polyhedral volumes ($\times 10^5$ km$^3$)          & 4.95140993      \\
                Masses estimation ($\times 10^{18}$ kg)            & 1.68347937      \\
                Dynamical polar ﬂattening: $J_2 (-C_{20})$         & 0.13047304      \\
                Dynamical equatorial ﬂattening $C_{22}$            & 0.03047707      \\
                Moments of inertia $I_{xx}/M$ (km$^2$)             & 802.929326      \\
                Moments of inertia $I_{yy}/M$ (km$^2$)             & 1096.55453      \\
                Moments of inertia $I_{zz}/M$ (km$^2$)             & 1263.99603      \\
                \multirow{3}{*}{}                                  & $a$ = 62.40235  \\
                Equivalent ellipsoid (km)                          & $b$ = 49.25370  \\
                                                                   & $c$ = 39.85875   \\
                \hline
            \end{tabular}
        }
    \end{table}

\section{Potential model, equilibrium points, and stability of asteroid (21) Lutetia}\label{potential_model}
    In this section, we begin by briefly describing the Potential Series Expansion Method (PSEM) \citet{Mota_2017} and present the results for asteroid (21) Lutetia, assuming a three-layered internal structure, as detailed in Table 2.

 \subsection{The Potential Series Expansion Method}\label{The Potential Series Expansion Method}  

In this study, we consider only the perturbation generated by the gravitational potential of the asteroid and establish a reference frame fixed to the body. The Potential Series Expansion Method (PSEM) is employed, together with the decomposition of the asteroid into tetrahedral elements, to model its gravitational potential. The center of mass and the principal axes of inertia of this polyhedron are aligned with the origin and axes of the coordinate system, respectively, as illustrated in Figure 1.
\begin{figure}[ht]   
         \includegraphics[width=1\linewidth]{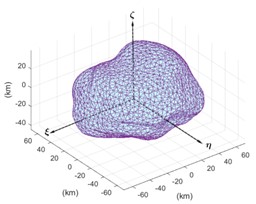}\\
        \caption{Asteroid (21) Lutetia decomposed into 2,962 tetrahedra.} \label{Asteroid Lutetia decomposed into 2962 tetrahedra}
    \end{figure} \\   
Let $Q_{k}$ be a generic tetrahedron of mass $M_k$  be defined by the vertices $V_{1k}$, $V_{2k}$, $V_{3k}$ and O, with the latter located at the origin of the coordinate system $\xi ,\eta ,\zeta $ , as shown in Figure 2.
  \begin{figure}[ht]   
         \includegraphics[width=1\linewidth]{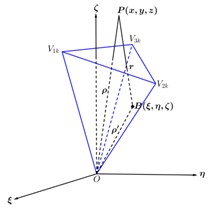}\\
        \caption{Tetrahedral $Q_{k}$ of vertices $V_{1k}$, $V_{2k}$, $V_{3k}$ and O, indicating the distances $\rho$, $\rho'$ and $r$ .} \label{tetraedro}
    \end{figure} 
        
  To calculate the potential of this tetrahedron, we approximate the total gravitational potential around the asteroid by summing the contributions of each tetrahedral element. This is done using the direct evaluation of the triple integral over for each tetrahedral volume, leading to Equation (1):
   \begin{eqnarray}\label{eq_potential_PSEM}
        U &=& G\sigma\sum_{k=1}^{n} \sum_{i=0}^{m} \iiint\limits_{Q_{k}}P_{i}(u)\frac{\rho'^{i}\rho^{i}}{\rho^{(2i+1)}}dV + \epsilon
    \end{eqnarray}
    \noindent where $G$ is the gravitational constant, $\sigma$ is the density of the asteroid, $Q_{k}$ is a tetrahedral element, $m$ is the degree of expansion, $n$ is the number of faces of the polyhedron, $\rho$ is the distance from a point outside the body to its center of mass, $\rho'$ is the distance of a point belonging to the body to its center of mass, as shown in Fig. \ref{tetraedro}, $P_{i}(u)$ is the Legendre polynomials and $\epsilon$ is the truncation error. For more details on this expression and the mathematical development, we refer the reader to \citet{Mota_2017}.\\
     
We define $U_i$ as the potential of degree $i=0,1,2,...,m$ that satisfies Equation (2)

\begin{eqnarray}\label{general_equation_of_potential}
       {{U}_{i}}=G\sigma \sum\limits_{k=1}^{n}{\iiint\limits_{{{Q}_{k}}}{{{P}_{i}}\left( u \right)\frac{{{{{\rho }'}}^{i}}{{\rho }^{i}}}{{{\rho }^{2i+1}}}\text{dV}}}
    \end{eqnarray}
and using PSEM, we conclude that the gravitational potential of the polyhedron $Q=\bigcup\limits_{k=1}^{n}{{{Q}_{k}}}$ can be approximated by Equation (3)
\begin{eqnarray}\label{sum_of_potentials}
       U={{U}_{0}}+{{U}_{1}}+{{U}_{2}}+...+{{U}_{m}}
    \end{eqnarray}
 where $U_0$ corresponds to the Keplerian potential, and the summation of the remaining terms represents the perturbation due to the non-central nature of the gravitational potential. 
 \subsection{The three-layered internal structure model}\label{The three-layered internal structure model}
 Our three-layered model is similar to the one discussed in \citet{2014Icar..240..118P} and \citet{2014GeoRL..41.1452K}. It corresponds to an equivalent volume diameter of 98.155 km, where: a crust with an average thickness of 18.404 km accounts for 75.59$\%$ of the total volume and has a density of 3.2 g.cm$^{-3}$, representing 71.06$\%$ of the total mass; the mantle is modeled with a thickness of 18.404 km (22.85$\%$ of the total volume) and a density of 3.8 g.cm$^{-3}$ (25.54$\%$ of the total mass); the core, based on the characteristics of iron meteorites, has a thickness of 12.27 km (1.56$\%$ of the total volume) and a density of 7.4 g.cm$^{-3}$ (3.4$\%$ of the total mass), ), as shown in Figure 3.
\begin{figure}[ht]
\centering
         \includegraphics[width=0.9\linewidth]{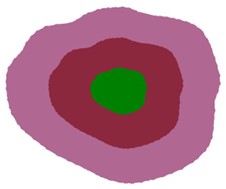}\\
        \caption{Three-layered structure of (21) Lutetia.} \label{Figure 3. Three-layered structure of (21) Lutetia.}
    \end{figure} \\ 
    The proposed three-layered internal structure for (21) Lutetia is summarized in Table 2. The layer thicknesses and densities are constrained by internal structure models of Vesta, as discussed in \citet{2014Icar..240..118P}, \citet{2014GeoRL..41.1452K}, and \citet{2011SSRv..163...77Z}. To preserve Lutetia’s total mass, we fix its average density at 3.4 g/cm$^{3}$. In other words, in the three-layered model, Lutetia’s gravitational distribution is adjusted to be denser at the center, while the average density remains the same as in the uniform-density model.\\
   
       \begin{table}[!htp]
        \caption{Three-layered structure of (21) Lutetia.} \label{table_Three-layered}
        \resizebox{0.5\textwidth}{!}{
            \begin{tabular}{lcccc}
                \hline
                Layer      & Thickness & Density & Volume & Mass \\ 
                           & (km)      &   (g/$cm^3$)     &   (\% of total)     &  (\% of total)     \\ \hline
                Core           & 12.270          &  7.40      &  1.56      &   3.40     \\ 
                Mantle          &  18.404         & 3.80      & 22.85       &  25.54     \\ 
                Crust           &  18.404         & 3.20      &  75.59      &  71.06      \\ 

                \hline
            \end{tabular}
        }
    \end{table}

  To analyze the effects of the internal structure described above on the external gravitational potential of (21) Lutetia, we use a polyhedral model consisting of 2,962 triangular faces and apply the Potential Series Expansion Method (PSEM) \citet{Mota_2017}. To illustrate the procedure, we consider a tetrahedral element and divide it into three volumes with parallel bases $A_1$$B_1$$C_1$, $A_k$$B_2$$C_2$ and $A_3$$B_3$$C_3$  , as shown in Figure 4. 
  \begin{figure}[ht]
  \centering
         \includegraphics[width=0.6\linewidth]{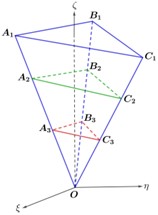}\\
        \caption{Division of a tetrahedral element into three layers.} \label{Figure 4. Three-layered structure of (21) Lutetia.}
    \end{figure} \\ 
To compute the total potential of the tetrahedral element $Q_{k}$  shown in Figure 4, we consider each layer with its corresponding density, according to Table 2. Defining ${{U}_{{{t}_{1}}{{\sigma }_{1}}}}$ as the potential of tetrahedron $A_1$$B_1$$C_1$$O$ with density ${{\sigma }_{1}}$, ${{U}_{{{t}_{2}}{{\sigma }_{1}}}}$ as the potential of tetrahedron $A_2$$B_2$$C_2$$O$ with density ${{\sigma }_{1}}$, ${{U}_{{{t}_{2}}{{\sigma }_{2}}}}$ as the potential of tetrahedron $A_2$$B_2$$C_2$$O$ with density ${{\sigma }_{2}}$, ${{U}_{{{t}_{3}}{{\sigma }_{2}}}}$ as the potential of tetrahedron $A_3$$B_3$$C_3$$O$ with density ${{\sigma }_{2}}$, and ${{U}_{{{t}_{3}}{{\sigma }_{3}}}}$ as the potential of tetrahedron $A_3$$B_3$$C_3$$O$ with density ${{\sigma }_{3}}$, then the total potential of the tetrahedral element $A_1$$B_1$$C_1$$O$, considering all three densities, is given by Equation (4).

     \begin{eqnarray}\label{eq_potential_layers}
       {{U}_{T}}=\sum\limits_{j=1}^{2}{\left( {{U}_{{{t}_{j}}{{\sigma }_{j}}}}-{{U}_{{{t}_{j+1}}{{\sigma }_{j}}}} \right)}+{{U}_{{{t}_{3}}{{\sigma }_{3}}}}
    \end{eqnarray}
    
Given the analytical expression of the potential of tetrahedron $A_1$$B_1$$C_1$$O$ and the proportionality ratios between similar tetrahedral, we have Equation (5),
 \begin{eqnarray}\label{ratio_between_volumes}
       \frac{{{V}_{{{A}_{2}}{{B}_{2}}{{C}_{2}}O}}}{{{V}_{{{A}_{1}}{{B}_{1}}{{C}_{1}}O}}}={{\left( \frac{5}{8} \right)}^{3}}\text{    and     }  \frac{{{V}_{{{A}_{3}}{{B}_{3}}{{C}_{3}}O}}}{{{V}_{{{A}_{1}}{{B}_{1}}{{C}_{1}}O}}}={{\left( \frac{2}{8} \right)}^{3}}
    \end{eqnarray}
  we can compute the potentials of tetrahedra $A_2$$B_2$$C_2$$O$ and $A_3$$B_3$$C_3$$O$ using the similarity ratios $5/8$  and $2/8$ , resulting in Equations (6) and (7),
  \begin{eqnarray}\label{eq_potential_2}
       {{U}_{{{t}_{2}}{{\sigma }_{i}}}}={{U}_{0}}+\left( \frac{5}{8} \right){{U}_{1}}+...+{{\left( \frac{5}{8} \right)}^{m}}{{U}_{m}}
    \end{eqnarray}
\begin{eqnarray}\label{eq_potential}
       {{U}_{{{t}_{3}}{{\sigma }_{j}}}}={{U}_{0}}+\left( \frac{2}{8} \right){{U}_{1}}+...+{{\left( \frac{2}{8} \right)}^{m}}{{U}_{m}}
    \end{eqnarray}
    
    where ${{\sigma }_{i}}$  and  ${{\sigma }_{j}}$ are the densities as specified in Table 2, $i=1,2$, $j=2,3$, and ${{U}_{1}},{{U}_{2}},...,{{U}_{m}}$ are given by Equation (3). 

 \subsection{Equilibrium points and stability considering a three-layered structure}\label{Equilibrium points and stability considering a three-layered structure}
In this subsection, we determine the coordinates of the equilibrium points and analyze their corresponding stability assuming that asteroid (21) Lutetia has a three-layered internal structure, as described in Table 2. We apply the Potential Series Expansion Method (PSEM) and compare the results obtained with the Mascon 8 gravity model \citet{aljbaae_2017} to validate our findings. For more details on the energy equation, we refer the reader to \citet{aljbaae_2017}, \citet{Jiang_2014} and \citet{Wang_2014}.\\

The positions of each equilibrium point obtained using the classical polyhedral method \citet{tsoulis_2001}, the Mascon 8 gravity model \citet{chanut_2015}, and PSEM \citet{Mota_2017} are presented in Table 3. Additionally, Table 4 shows the relative position vector errors for the equilibrium points computed using Mascon 8 and PSEM.\\
   \begin{table}[!htp]
       \caption{Locations of equilibrium points of (21) Lutetia and their 
Jacobi constant $C$ (using the shape model with 2,962 faces)} \label{equilibrium_points}
       \begin{center}            
       \resizebox{0.48\textwidth}{!}{
           \begin{tabular}{|ccccc|}
               \hline
                 & x (km) & y (km) & z (km) &  $C$($km^2$ $s^{-2}$)\\\hline
                 \multicolumn{5}{|c|}{Polyhedral, uniform density (\citet{tsoulis_2001})}{}\\\hline
    $E1$   &  137.10784172  &  8.44279347  &  0.08555291  &  -0.12634256 $\times 10^{-2}$    \\\hline
    $E2$   & -138.19144378  &  6.56551358  &  0.04185644  &  -0.12679936  $\times 10^{-2}$    \\\hline
    $E3$   &  -8.70389476   &  134.01690523 &  0.03696436 &  -0.12441019   $\times 10^{-2}$   \\\hline
    $E4$   &  -14.61831274  & -134.06107222 &  0.08749509 &  -0.12467120   $\times 10^{-2}$      \\\hline
            \multicolumn{5}{|c|}{Mascon 8 (\citet{aljbaae_2017}), three-layered structure}{}\\\hline
    $E1$   &  136.99825452  & 8.32304070  &  0.08727020 & -0.12626501   $\times 10^{-2}$     \\\hline
    $E2$   & -138.01547466  & 6.50802298  &  0.04322724 & -0.12669215   $\times 10^{-2}$      \\\hline
    $E3$   & -8.61381286    & 134.06254558 & 0.03698190 &  -0.12443281    $\times 10^{-2}$    \\\hline
    $E4$   & -14.33458820   & -134.09482376 & 0.08998671 & -0.12467656   $\times 10^{-2}$   \\\hline
                \multicolumn{5}{|c|}{PSEM (\citet{Mota_2017}), three-layered structure }{}\\\hline
     $E1$  &  137.06436904 & 8.37985602 & 0.080746509 & -0.12636679  $\times 10^{-2}$      \\\hline
     $E2$  & -138.10149787 & 6.53841465 &  0.041750504 & -0.12680234  $\times 10^{-2}$     \\\hline
     $E3$  & -8.660766977  & 134.09774745 & 0.035437310 & -0.12451363  $\times 10^{-2}$     \\\hline
     $E4$  & -14.47909699  & -134.12679947 & 0.083302742 &   -0.12476182 $\times 10^{-2}$      \\\hline
           \end{tabular}       
           }
       \end{center}
   \end{table}
The relative position vector errors of the equilibrium points obtained using Mascon 8 and PSEM are presented in Table 4, confirming that PSEM provides reasonable accuracy.\\
\begin{table}[!htp]
\centering
       \caption{. Relative position vector errors (\%) of the equilibrium points (E) obtained using Mascon 8 and PSEM}
       \resizebox{0.18\textwidth}{!}{
           \begin{tabular}{cc}
               \hline
                &  ER (\%) \\\hline
            $E1$    &   0.0343956       \\\hline
            $E2$    &   0.0658695       \\\hline
            $E3$    &   0.0579939       \\\hline
            $E4$    &   0.0373167       \\\hline
           \end{tabular}
       }
   \end{table}
Additionally, Table 5 presents the execution time required to compute the gravitational potential at 1,002,000 points near Lutetia using a Pentium 3.60 GHz CPU. The results confirm that our method significantly reduces computational cost compared to the classical polyhedral method, while maintaining an acceptable level of accuracy.\\

    \begin{table}[!htp]
    \centering
        \caption{Execution time for calculating the gravitational potential on a 1,002,000 point close to Lutetia using a Pentium 3.60GHz CPU.} \label{table_execution_time}
        \resizebox{0.4\textwidth}{!}{
            \begin{tabular}{ll}
                \hline
                \citet{tsoulis_2001}  & This work \\
                \hline
                111m49.747s & 0m25.004s \\
                \hline
            \end{tabular}
        }
    \end{table}
  Based on the equilibrium point locations obtained for asteroid (21) Lutetia using PSEM (as shown in Table 3), we determine the eigenvalues associated with each point, which are, as presented in Table 6.  
    \begin{table}[!htp]
        \caption{Eigenvalues of the equilibrium points around the asteroid (21) Lutetia.} \label{Eigenvalues}
        \resizebox{0.5\textwidth}{!}{
            \begin{tabular}{|l|l|l|l|l|}
                \hline
                  $\times 10^{-3}$& $E_1$ & $E_2$ &  $E_3$ & $E_4$\\
                \hline
                $\lambda_1$ & 0.0675 & 0.0941 & 0.0958$i$ & 0.0746$i$\\
                \hline
                $\lambda_2$  & -0.0675 & -0.0941 & -0.0958$i$ & -0.0746$i$\\
                \hline
                $\lambda_3$  & 0.218$i$ & 0.223$i$ & 0.189$i$ & 0.196$i$ \\
                \hline
                $\lambda_4$  & -0.218$i$ & -0.223$i$ & -0.189$i$ & -0.196$i$\\
                \hline
                $\lambda_5$  & 0.220$i$ & 0.225$i$ & 0.215$i$ & 0.217$i$\\
                \hline
                $\lambda_6$  & -0.220$i$ & -0.225$i$ & -0.215$i$ & -0.217$i$\\                \hline
            \end{tabular}
        }
    \end{table}
 From Table 6, we observe that equilibrium points E1 and E2 are unstable, while points E3 and E4 are linearly stable. According to the classification by \citet{Jiang_2014} and \citet{Wang_2014}, equilibrium points E1 and E2 exhibit one pair of real eigenvalues and two pairs of purely imaginary eigenvalues, characterizing them as saddle-type equilibrium points (Case 2). These points support two families of periodic orbits and one family of quasi-periodic orbits in their vicinity. In contrast, equilibrium points E3 and E4 have three pairs of purely imaginary eigenvalues, classifying them as center-type equilibrium points (Case 1), indicating the presence of three families of periodic orbits around each. These findings reinforce the validity of our three-layered model in representing the gravitational environment of (21) Lutetia and demonstrate the effectiveness of PSEM in accurately locating equilibrium points and evaluating their stability.\\

 Figure 5 presents the projection of the zero-velocity surface onto the xy-plane, along with the locations of the equilibrium points computed using PSEM (three-layered structure) for asteroid (21) Lutetia.
     \begin{figure}[ht] 
     \centering
        \includegraphics[width=0.99\linewidth]{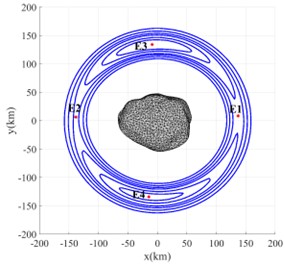}\\
       \caption{Zero-velocity curves and equilibrium points near asteroid (21) Lutetia in the xy-plane, obtained using the 2,962-face polyhedral shape model and a three-layered internal structure.} \label{fig_eq_points}
    \end{figure}

\section{Conclusion}\label{conclusion}

This study applied the Potential Series Expansion Method (PSEM) \citet{Mota_2017} to investigate the effects of internal structure on the gravitational potential modeling of asteroid (21) Lutetia, considering a three-layered composition with distinct densities. Initially, the physical properties of the asteroid were analyzed under the assumption of uniform density. Subsequently, by incorporating a three-layered internal structure with varying densities, PSEM was applied to derive a a model of the external gravitational field. Based on this model, we determined the coordinates of the equilibrium points and analyzed their corresponding stability, comparing the results with those obtained using the classical polyhedral method and the Mascon 8 model.
The PSEM proved to be an efficient method, significantly reducing computational processing time while maintaining high accuracy in modeling irregularly shaped bodies. The results indicate that PSEM is a valuable tool for modeling the gravitational potential of asteroids with multilayered internal structures, offering a substantial reduction in the computational cost of simulating asteroid orbits.

\section*{Acknowledgements}\label{Acknowledgements}
    The authors wish to express their appreciation for the support provided by: grant 309089/2021-2 from the National Council for Scientific and Technological Development (CNPq) and the Coordination for the Improvement of Higher Education Personnel (CAPES).  
\section*{Data Availability}\label{Data Availability}
All the data and the code used in this work will be available
from the first author upon reasonable request.

\section*{Conflict of interest}\label{Conflict of interest}
The authors declare no conflicts of interest. 

%
%

  \bibliographystyle{abbrvnat}
  \bibliography{mybibliography}

\end{document}